\documentclass[%
 aip,jcp,
 amsmath,amssymb,
 reprint,
]{revtex4-1}

\usepackage{graphicx}
\usepackage{dcolumn}
\usepackage{bm}

\usepackage[utf8]{inputenc}
\usepackage[T1]{fontenc}
\usepackage{mathptmx}
\usepackage{listings}
\usepackage{rotating} 

\usepackage{listings} 

\usepackage{xcolor}

\definecolor{codegreen}{rgb}{0,0.6,0}
\definecolor{codegray}{rgb}{0.5,0.5,0.5}
\definecolor{codepurple}{rgb}{0.58,0,0.82}
\definecolor{backcolour}{rgb}{0.95,0.95,0.92}
\lstdefinestyle{mystyle}{
  backgroundcolor=\color{backcolour},   commentstyle=\color{codegreen},
  keywordstyle=\color{magenta},
  numberstyle=\tiny\color{codegray},
  stringstyle=\color{codepurple},
  basicstyle=\ttfamily\scriptsize,
  breakatwhitespace=false,         
  breaklines=true,                 
  captionpos=b,                    
  keepspaces=true,                 
  showspaces=false,                
  showstringspaces=false,
  showtabs=false,
  xleftmargin=0.02\textwidth,
  rulecolor=\color[RGB]{200,200,200},
  frame=bt,
  framextopmargin=2pt,
  framexbottommargin=2pt,
  framexleftmargin=10pt,
  tabsize=2
}
\newcommand{\pq}{\texttt{p$^\dagger$q~}}
\newcommand{\np}{\textsc{NumPy}}
\newcommand{\app}[1]{\hyperref[app:#1]{Appendix~\ref*{app:#1}}}

\begin{document}

\author{Nicholas C. Rubin}
\affiliation{
             Google Research, 
             Mountain View, CA, USA}
\email{nickrubin@google.com}

\author{A. Eugene DePrince III}
\affiliation{
             Department of Chemistry and Biochemistry,
             Florida State University,
             Tallahassee, FL 32306-4390}
\email{adeprince@fsu.edu}

\title{\pq: A tool for prototyping many-body methods for quantum chemistry}

\begin{abstract}

\pq is a C++ accelerated Python library designed to generate equations for many-body quantum chemistry methods and to realize proof-of-concept implementations of these equations for rapid prototyping. Central to this library is a simple interface to define strings of second-quantized creation and annihilation operators and to bring these strings to normal order with respect to either the true vacuum state or the Fermi vacuum. Tensor contractions over fully-contracted strings can then be evaluated using standard Python functions ({\em e.g.}, \np's einsum). Given one- and two-electron integrals, these features allow for the rapid implementation and assessment of a wide array of many-body quantum chemistry methods. 

\end{abstract}

\maketitle

\section{Introduction}

The development of many-body methods for quantum chemistry applications carries several technical challenges. Deriving working equations by hand can be tedious and error prone, and the realization of even a proof-of-concept implementation that exhibits reasonable performance can be difficult, particularly for recent entrants to the field. Consequently, there is a long history\cite{Hirata06_2} of applying computers for automated generation, manipulation, and implementation of the working equations of quantum chemistry methods. 
Automation strategies of varying sophistication have been applied in {\color{black}a variety of contexts, such as} many-body perturbation theory (MBPT), \cite{Wong73_1,Paldus73_9, Bartlett85_151, Hirao85_8} coupled-cluster (CC) theory\cite{Bartlett82_1910,Cizek90_831,Schaefer91_1, Bartlett91_387, Paldus96_1210, Schaefer97_7943, Harris99_593, Jeziorski99_1857, Lotrich00_494, Lotrich00_4549, Bartlett00_216, Surjan00_1359, Olsen00_7140, Surjan01_2945, Gauss02_7872, Sherrill04_3374,Kohn09_131101,Kohn10_174117,Kohn10_174118,Gauss13_2639,Neese17_1853} ({\color{black}including response properties\cite{Kohn09_124118,Kohn09_104104,Kohn18_064101} and analytic gradients\cite{Gauss04_6841,Nooijen05_1,Gauss14_104102}),} multireference (MR) approaches,\cite{Paldus94_8812, Lotrich01_253,Nooijen02_656,Adamowicz05_024108,Kohn11_204111,Kohn12_204107,Kohn12_131103,Shiozaki15_051103,Shiozaki16_3781,Kohn18_693,Kohn18_064101} {\color{black} and reduced density matrix (RDM) methods.\cite{Hazra03_4832}}
Among the modern examples of automated code generation for quantum chemistry, the tensor contraction engine (TCE)\cite{Hirata03_9887}, the Symbolic Manipulation Interpreter for Theoretical cHemistry (\textsc{smith}3)\cite{SMITH3}, {\color{black} the General Contraction Code (GeCCo),\cite{Tew08_201103}} and the computational science that these tools enable, stand out as prime examples of the utility of this approach to code development.

The primary utility of automation is clear: it eliminates human error. Indeed, documented examples\cite{Malmqvist17_2077} wherein machine-generated algorithms have revealed bugs in mature electronic structure packages highlight the power of such tools. 
Beyond this obvious benefit, a well-designed interface to a fermion algebra engine also has the potential to accelerate the development and assessment of novel quantum chemistry methods. Consider, for example, the large number of existing many-body approaches that can be realized via selective elimination or modification of diagrams appearing in the CC with single and double excitations (CCSD)\cite{Bartlett82_1910} equations ({\em e.g.} quadratic configuration interaction (QCI),\cite{Raghavachari87_5968}  approximate second-order CC [CC2],\cite{Jorgensen95_409} nCC\cite{Musial06_204105} models,  parametrized CCSD [pCCSD]\cite{Nooijen10_184109,Nooijen12_064101,Bartlett17_144104}, etc.). The development of pCCSD, for example, was motivated by the search for scaling parameters such that a CCSD-like method could mimic the effects of connected triple excitations while remaining exact for two-electron systems. Given a sufficiently flexible code generation platform, more general parametrized CC theories involving larger parameter spaces could be explored in a completely automated fashion. The utility of automated approaches to method development becomes even more apparent when considering modifications to or parametrizations of higher-order CC methods, {\em e.g.}, CCSD plus triple excitations (CCSDT)\cite{Bartlett87_7041} models beyond familiar iterative\cite{Helgaker97_1808} and non-iterative\cite{HeadGordon89_479} approximations. Our present efforts are partially motivated by the potential utility of a user-friendly framework for just such explorations.

We have developed a C++ accelerated Python library,\cite{pq_2020}  \texttt{p$^{\dagger}$q}, for 
symbolic fermion algebra that can be used to derive and implement equations for general many-body quantum chemistry methods through a simple and intuitive user interface. At the core of this library is a C++ engine for manipulating strings of fermionic creation and annihilation operators in order to bring them to normal order with respect to either the Fermi vacuum or the true vacuum state. Users can interface with this engine through simple Python calls that support a variety of built-in second-quantized operator types ({\em e.g.}, cluster amplitudes, the fluctuation potential, etc.) and operations ({\em e.g.}, commutators, similarity transformations, etc.). Once a list of strings has been defined and brought to normal order, the code collects like terms to deliver compact lists of fully-contracted strings that can either be printed for subsequent use/analysis by the user or translated by \pq into tensor contractions that are evaluated using \np's einsum. Given one- and two-electron integrals obtained externally (from OpenFermion\cite{Babbush20_034014}, for example), these automatically generated contractions can then be easily incorporated by the user into appropriate kernels to realize a functioning quantum chemistry code. 

In this work, we provide an overview of the basic capabilities of \pq and demonstrate its use as a code generation platform. In Sec.~\ref{SEC:TOOLS}, we explore a typical use case in which normal-order is defined relative to the Fermi vacuum, providing a walk-through of the generation of spin-orbital-basis equations for determining cluster amplitudes, de-excitation amplitudes, and unrelaxed RDMs at the CCSD level of theory. Additional use cases involving normal-order with respect to the true vacuum are also considered. The procedure by which the equations generated by \pq can be converted into executable code via \np~einsum contractions is discussed in Sec.~\ref{SEC:CODE_GENERATION}. Some concluding remarks can be found in Sec.~\ref{SEC:CONCLUSIONS}.

\section{Fermion Algebra}

\label{SEC:TOOLS}

\vspace{-0.2cm}
\subsection{Built-in operator types and operations}
\vspace{-0.2cm}

\pq supports several built-in operator types, including CC amplitudes, CC de-excitation amplitudes (or left-hand equation of motion [EOM] CC\cite{Bartlett93_7029,Bartlett12_126,Krylov08_433} amplitudes), right-hand EOM-CC amplitudes, and general one- and two-body operators. A complete list of these operators and their definitions are provided in Table \ref{tab:operators}. Note that we use standard notation for labeling orbital spaces, {\em i.e.}, $i$, $j$, $k$, $l$, $m$, and $n$ refer to spin-orbitals that are occupied in the reference configuration; $a$, $b$, $c$, $d$, $e$, and $f$ are unoccupied spin-orbitals; and general spin-orbitals are denoted by labels $p$, $q$, $r$, and $s$. The library also supports enough basic operator manipulation to compactly define equations for many-body methods such as CCSD (see Table \ref{tab:functions}). Before delving into such a complex case, we first introduce some important features of \pq through a much simpler example. 

\begin{table*}[!htpb]
    \centering
    \caption{{\color{black}Operator types supported by \pq.} }
    \label{tab:operators}
    \begin{tabular}{lll}
    \hline\hline
            operator symbol~~~~ & operator definition & operator description\\
        \hline
        \texttt{t1} & $t_i^a \hat{a}^\dagger_a \hat{a}_i$ & singles cluster amplitudes \\
        \texttt{t2} & $\frac{1}{{\color{black}(2!)^2}} t_{ij}^{ab} \hat{a}^\dagger_a \hat{a}^\dagger_b \hat{a}_j \hat{a}_i$ & doubles cluster amplitudes \\
        \texttt{t3} & $\frac{1}{{\color{black}(3!)^2}} t_{ijk}^{abc} \hat{a}^\dagger_a \hat{a}^\dagger_b \hat{a}^\dagger_c \hat{a}_k \hat{a}_j \hat{a}_i$ & triples cluster amplitudes \\
        {\color{black}\texttt{t4}} & {\color{black}$\frac{1}{(4!)^2} t_{ijkl}^{abcd} \hat{a}^\dagger_a \hat{a}^\dagger_b \hat{a}^\dagger_c \hat{a}^\dagger_d \hat{a}_l \hat{a}_k \hat{a}_j \hat{a}_i$} & {\color{black}quadruples cluster amplitudes} \\
        \texttt{r1} & $r_i^a \hat{a}^\dagger_a \hat{a}_i$ & EOM-CC right-hand singles amplitudes  \\
        \texttt{r2} & $\frac{1}{{\color{black}(2!)^2}} r_{ij}^{ab} \hat{a}^\dagger_a \hat{a}^\dagger_b \hat{a}_j \hat{a}_i$ & EOM-CC right-hand doubles amplitudes  \\
        {\color{black}\texttt{r3}} & {\color{black}$\frac{1}{(3!)^2} r_{ijk}^{abc} \hat{a}^\dagger_a \hat{a}^\dagger_b \hat{a}^\dagger_c \hat{a}_k \hat{a}_j \hat{a}_i$} & {\color{black}EOM-CC right-hand triples amplitudes} \\
        {\color{black}\texttt{r4}} & {\color{black}$\frac{1}{(4!)^2} r_{ijkl}^{abcd} \hat{a}^\dagger_a \hat{a}^\dagger_b \hat{a}^\dagger_c \hat{a}^\dagger_d \hat{a}_l \hat{a}_k \hat{a}_j \hat{a}_i$} & {\color{black}EOM-CC right-hand quadruples amplitudes} \\
        \texttt{l0}~~~ & $l_0$ & EOM-CC left-hand reference amplitude  \\
        \texttt{l1}~~~ & $l^i_a \hat{a}^\dagger_i \hat{a}_a$ & EOM-CC left-hand singles amplitudes \\
        \texttt{l2}~~~ & $\frac{1}{{\color{black}(2!)^2}} l^{ij}_{ab} \hat{a}^\dagger_i \hat{a}^\dagger_j \hat{a}_b \hat{a}_a$ & EOM-CC left-hand doubles amplitudes \\
        {\color{black}\texttt{l3}} & {\color{black}$\frac{1}{(3!)^2} l^{ijk}_{abc} \hat{a}^\dagger_i \hat{a}^\dagger_j \hat{a}^\dagger_k \hat{a}_c \hat{a}_b \hat{a}_a$} & {\color{black}EOM-CC left-hand triples amplitudes} \\
        {\color{black}\texttt{l4}} & {\color{black}$\frac{1}{(4!)^2} l^{ijkl}_{abcd} \hat{a}^\dagger_i \hat{a}^\dagger_j \hat{a}^\dagger_k \hat{a}^\dagger_l \hat{a}_d \hat{a}_c \hat{a}_b \hat{a}_a$} & {\color{black}EOM-CC left-hand quadruples amplitudes} \\
        \texttt{1}     & $1$ & unit operator \\
        \texttt{h}     & $\sum_{pq} h_{pq} \hat{a}^\dagger_p \hat{a}_q$ & general one-body operator \\
        \texttt{g}     & $\sum_{pqrs} g_{pqrs} \hat{a}^\dagger_p \hat{a}^\dagger_q \hat{a}_s \hat{a}_r$ & general two-body operator \\
        \texttt{f}       & $\sum_{pq} (h_{pq} + \sum_i \langle pi||qi \rangle) \hat{a}^\dagger_p \hat{a}_q$ & Fock operator \\
        \texttt{v}       & $\frac{1}{4} \sum_{pqrs} \langle pq||rs \rangle \hat{a}^\dagger_p \hat{a}^\dagger_q \hat{a}_s \hat{a}_r - \sum_{pqi} \langle pi||qi \rangle \hat{a}^\dagger_p \hat{a}_q$~~~~ & fluctuation potential operator \\
        \texttt{e1(p,q)} & $\hat{a}^\dagger_p \hat{a}_q$ & one-body transition operator \\
        \texttt{e2(p,q,r,s)} & $\hat{a}^\dagger_p \hat{a}^\dagger_q \hat{a}_r \hat{a}_s$ & two-body transition operator \\
        \texttt{e3(p,q,r,s,t,u)} & $\hat{a}^\dagger_p \hat{a}^\dagger_q \hat{a}^\dagger_r \hat{a}_s \hat{a}_t \hat{a}_u$ & three-body transition operator \\
        \texttt{e4(p,q,r,s,t,u,v,w)} & $\hat{a}^\dagger_p \hat{a}^\dagger_q \hat{a}^\dagger_r \hat{a}^\dagger_s \hat{a}_t \hat{a}_u \hat{a}_v \hat{a}_w $ & four-body transition operator \\
        \hline\hline
    \end{tabular}
\vspace{-0.4cm}
\end{table*}

\begin{sidewaystable}
    \centering
    \caption{{\color{black}\pq function calls and descriptions. The letters a, b, etc. refer to operators tabulated in Table \ref{tab:operators}, and the notations $a+b$ or $ab$ refer to sums or products of operators a and b, respectively. The notation $num \times$ indicates that the operator products/sums are scaled by a numerical factor, num.}}
    \label{tab:functions}
    \begin{tabular}{ll}
    \hline\hline
        function call & description \\
        \hline
        \texttt{pq = pq\_helper(vacuum\_type)} ~~\#~~ vacuum\_type = 'fermi'/'true'  & initialize operator class with vacuum type \\
        \texttt{pq.set\_left\_operators([a,b,...])} & define bra state as $\langle \psi_0| (a + b + ...)$\\
        \texttt{pq.set\_right\_operators([a,b,...])} & define ket state as $(a + b + ...)|\psi_0\rangle$\\
        \texttt{pq.add\_operator\_product(num, [a,b,...])} & $num\times ab$... \\
        \texttt{pq.add\_commutator(num, [a,b,...],[c,d,...])}  & $num\times[ab...,cd...]$\\
        \texttt{pq.add\_double\_commutator(num, [a,b,...],[c,d,...],[e,f,...])}  & $num\times[[ab...,cd...],ef...]$\\
        \texttt{pq.add\_triple\_commutator(num, [a,b,...],[c,d,...],[e,f,...],[g,h,...])}  & $num\times[[[ab...,cd...],ef...],gh...]$ \\
        \texttt{pq.add\_quadruple\_commutator(num, [a,b,...],[c,d,...],[e,f,...],[g,h,...],[i,j...])}  & $num\times[[[[ab...,cd...],ef...],gh...],ij...]$ \\
        \texttt{pq.add\_st\_operator(num,[a,b,...],[c,d,...])} & $num\times e^{(-c-d-...)} (ab...) e^{(c+d+...)}$ \\
        \texttt{pq.print()} & print current list of strings \\
        \texttt{strings = pq.strings()} & return current list of strings \\
        \texttt{pq.simplify()} & {\color{black}simplify the current list of strings by canceling/combining terms}\\
        \texttt{pq.print\_fully\_contracted()} & print fully contracted strings \\
        \texttt{strings = pq.fully\_contracted\_strings()} & return list of fully contracted strings\\
        \texttt{pq.clear()} & clear current list of strings; reset bra/ket\\

            \hline\hline
    \end{tabular}
\end{sidewaystable}

The following code will evaluate the expectation value of the Hamiltonian with respect to a single reference configuration, which will serve as the vacuum state. 
\begin{lstlisting}
import pdaggerq
pq = pdaggerq.pq_helper('fermi')
pq.add_operator_product(1.0,['f'])
pq.add_operator_product(1.0,['v'])
pq.simplify()
terms = pq.fully_contracted_strings()
for my_term in terms:
    print(my_term)
pq.clear()
\end{lstlisting}
First, the argument passed to \texttt{pq\_helper} indicates that normal order is defined with respect to the Fermi vacuum. Second, the function \texttt{add\_operator\_product} adds the Fock and fluctuation potential operators the current list of operator strings (see Table \ref{tab:operators} for the definitions of these operators). After each of these calls, the relevant string is automatically brought to normal order. In a similar way, all functions in Table \ref{tab:functions} labeled \texttt{add\_xxx} automatically bring the argument strings to normal order when they are invoked. The \texttt{simplify} function then cleans up the resulting list of normal-ordered strings by removing delta functions (arising from the application of anticommutation relations $\{\hat{a}^\dagger_p,\hat{a}_q\} = \delta_{pq}$) and consolidating or canceling like terms. The user can then request a list of fully contracted strings. Lastly, the \texttt{clear} function clears the current list of strings so that the \texttt{pq} object can be used to define additional operator products, etc. This simple code produces the output
\begin{lstlisting}
['+1.000000', 'f(i,i)']
['-0.500000', '<i,j||i,j>']
\end{lstlisting}
where repeated labels imply sums, which, in this case, are carried out over occupied orbitals.

We note here that the string argument that is passed to \texttt{add\_operator\_product} is actually a list of strings that are interpreted as a product of operators by \pq. In this way, equations for essentially any many-body method can be developed via suitable calls to this function. Such a strategy will become cumbersome, though, for theories more complex than low-order perturbation theory, for example. In the following subsection, we introduce additional functions that allow for complex operator manipulations with simple and intuitive calls. 

\vspace{-0.2cm}
\subsection{Unrelaxed CCSD RDMs}
\vspace{-0.2cm}

The primary strength of \pq is the ease with which one can generate equations for sophisticated many-body methods. Here, we consider coupled-cluster (CC) theory,\cite{Cizek66_4256,Paldus71_359,Bartlett09_book,Musial07_291}  with a focus on the case of CC with single and double excitations (CCSD).\cite{Bartlett82_1910} Our goal is to explore enough functionality in \pq to realize a spin-orbital-basis implementation of CCSD unrelaxed one- and two-particle RDMs. {\color{black}Solvers for higher-order approaches ({\em e.g.} CCSD(T)\cite{HeadGordon89_479}, CC3,\cite{Helgaker97_1808} CCSDT,\cite{Bartlett87_7041} and CCSDTQ\cite{Adamowicz91_6645}) can be found on GitHub.\cite{pq_2020}}

We begin by defining the CCSD Lagrangian\cite{Schaefer90_4924}
\begin{equation}
    \label{EQN:LAGRANGIAN}
    L({\bm l},{\bm t}) = \langle \psi_0 | (1+\hat{\Lambda})e^{-\hat{T}} \hat{H} e^{\hat{T}} | \psi_0 \rangle
\end{equation}
where the cluster operator, $\hat{T}$, and the de-excitation operator, $\hat{\Lambda}$, are defined as
\begin{equation}
    \hat{T} = \sum_{ia} t_i^a \hat{a}^\dagger_a \hat{a}_i + \frac{1}{4} \sum_{ijab} t_{ij}^{ab}\hat{a}^\dagger_a \hat{a}^\dagger_b \hat{a}_j\hat{a}_i
\end{equation}
and
\begin{equation}
    \hat{\Lambda} = \sum_{ia} {\color{black}l^i_a} \hat{a}^\dagger_i \hat{a}_a + \frac{1}{4} \sum_{ijab} l^{ij}_{ab}\hat{a}^\dagger_i \hat{a}^\dagger_j \hat{a}_b\hat{a}_a
\end{equation}
respectively. Here, $\hat{a}^\dagger$ and $\hat{a}$ represent fermionic creation and annihilation operators, respectively, and $|\psi_0\rangle$ represents a reference determinant of {\color{black}orthonormal} spin-orbitals. The amplitudes ${\bm t}$ and ${\bm l}$ may be be determined by making Eq.~\ref{EQN:LAGRANGIAN} stationary with respect to their variations, {\em i.e.}, 
\begin{eqnarray}
\label{EQN:T_SINGLES}
    \frac{\partial L}{\partial l_e^m} = 0 & = & \langle \psi_0 | \hat{a}^\dagger_m \hat{a}_e e^{-\hat{T}} \hat{H} e^{\hat{T}} | \psi_0 \rangle,\\
\label{EQN:T_DOUBLES}
    \frac{\partial L}{\partial l_{ef}^{mn}} = 0 & = & \langle \psi_0 | \hat{a}^\dagger_m \hat{a}^\dagger_n \hat{a}_f \hat{a}_e e^{-\hat{T}} \hat{H} e^{\hat{T}} | \psi_0 \rangle,\\
\label{EQN:L_SINGLES}
    \frac{\partial L}{\partial t^e_m} = 0 & = & \langle \psi_0 | e^{-\hat{T}} \hat{H}  \hat{a}^\dagger_e \hat{a}_m e^{\hat{T}}| \psi_0 \rangle \nonumber \\ 
    & +&  \langle \psi_0 | \hat{\Lambda} e^{-\hat{T}} [\hat{H}, \hat{a}^\dagger_e \hat{a}_m] e^{\hat{T}} | \psi_0 \rangle,
\end{eqnarray}
and
\begin{align}\label{EQN:L_DOUBLES}
\frac{\partial L}{\partial t^{ef}_{mn}} = 0 = \langle \psi_0 | e^{-\hat{T}} \hat{H} \hat{a}^\dagger_e\hat{a}^\dagger_f \hat{a}_n \hat{a}_m e^{\hat{T}}| \psi_0 \rangle \nonumber \\ + \langle \psi_0 | \hat{\Lambda} e^{-\hat{T}} [\hat{H}, \hat{a}^\dagger_e \hat{a}^\dagger_f \hat{a}_n\hat{a}_m] e^{\hat{T}} | \psi_0 \rangle .
\end{align}
Equations \ref{EQN:T_SINGLES} and \ref{EQN:T_DOUBLES} are solved for the singles and doubles cluster amplitudes, respectively, while the singles and doubles de-excitation amplitudes can be determined via Eqs.~\ref{EQN:L_SINGLES} and \ref{EQN:L_DOUBLES}, respectively.

Python scripts for evaluating the singles and doubles cluster and de-excitation amplitudes according to Eqs.~\ref{EQN:T_SINGLES}--\ref{EQN:L_DOUBLES} are provided in Appendices \ref{app:ccsd_equation_gen} and \ref{app:lambda_equations}. Here, we  highlight the essential details necessary for understanding these examples. Consider first the following code that generates the singles equations defined by Eq.~\ref{EQN:T_SINGLES}
\begin{lstlisting}
# 0 = < 0 | m* e e(-T) H e(T) | 0> 
pq.set_left_operators(['e1(m,e)'])
pq.add_st_operator(1.0,['f'],['t1','t2'])
pq.add_st_operator(1.0,['v'],['t1','t2'])
pq.simplify()
t1_terms = pq.fully_contracted_strings()
for my_term in t1_terms:
    print(my_term)
pq.clear()
\end{lstlisting}
A bra state is defined by the \texttt{set\_left\_operators} function, after which the similarity-transformed Fock and fluctuation potential operators are added via the \texttt{add\_st\_operator} function (see Tables \ref{tab:operators} and \ref{tab:functions} for details). The latter functions evaluate the similarity transformation according to the Baker–Campbell–Hausdorff expansion, including up to four nested commutators. Obviously, many of these commutators will evaluate to zero ({\em e.g.} quadruple commutators involving \texttt{t2}), but the work to evaluate these zeros could be avoided through the use of lower-level commands specifying the non-zero commutators, double commutators, etc. (see Table \ref{tab:functions}). Equations for the doubles cluster amplitudes could be  obtained in a similar way; one need only change the bra state via \texttt{pq.set\_left\_operators(['e2(m,n,f,e)'])} (see Appendix \ref{app:ccsd_equation_gen}).

Given optimal cluster amplitudes, CCSD de-excitation amplitudes are defined via the solution of Eqs.~\ref{EQN:L_SINGLES} and \ref{EQN:L_DOUBLES}. For example, the following code will generate the singles de-excitation equations defined by Eq.~\ref{EQN:L_SINGLES} \\
\begin{lstlisting}
#  <0| e(-T) H e*m e(T)|0>
pq.set_left_operators(['1'])
pq.add_st_operator(1.0,['f','e1(e,m)'],['t1','t2'])
pq.add_st_operator(1.0,['v','e1(e,m)'],['t1','t2'])

# <0| L e(-T) [H,e*m] e(T)|0>
pq.set_left_operators(['l1','l2'])
pq.add_st_operator( 1.0,['f','e1(e,m)'],['t1','t2'])
pq.add_st_operator( 1.0,['v','e1(e,m)'],['t1','t2'])
pq.add_st_operator(-1.0,['e1(e,m)','f'],['t1','t2'])
pq.add_st_operator(-1.0,['e1(e,m)','v'],['t1','t2'])
pq.simplify()
l1_terms = pq.fully_contracted_strings()
for my_term in l1_terms:
    print(my_term)
pq.clear()
\end{lstlisting}
As compared to the code snippet for the singles cluster amplitudes, additional features to note here are (1) the bra state can be set such that it is defined by a sum of operators (in this case, \texttt{l1}+\texttt{l2}), and (2) the similarity transformation function can transform products of operators ({\em e.g.}, \texttt{v}$~\times$~\texttt{e1(e,m)}). As above, equations for the doubles de-excitation amplitudes could be  obtained in a similar way; one need only  swap all occurrences of \texttt{e1(e,m)} for \texttt{e2(m,n,f,e)}; a complete script is provided in Appendix \ref{app:lambda_equations}.

Now, given optimal cluster and de-excitation amplitudes, the definition of one- and two-particle RDMs within \pq is trivial. The following code will evaluate equations for the occupied-occupied block of the one-particle RDM\\
\begin{lstlisting}
# D(mn) = <0|(1 + l1 + l2) e(-T) m*n e(T) |0> 
pq.set_left_operators(['1','l1','l2'])
pq.add_st_operator(1.0,['e1(m,n)'],['t1','t2'])
pq.simplify()
d1_terms = pq.fully_contracted_strings()
for my_term in d1_terms:
    print(my_term)
pq.clear()
\end{lstlisting}
Additional blocks of the one-particle RDM can be generated through suitable choices of the labels in the \texttt{e1} operator, and blocks of the two-particle RDM can be defined via the same code, with \texttt{e1(m,n)} swapped in favor of two-body excitation operators, \texttt{e2(m,n,f,e)}, etc. Complete scripts fully defining all blocks of the CCSD unrelaxed one- and two-particle RDMs can be found in the Appendices \ref{app:1rdm} and \ref{app:2rdm}, respectively.

\vspace{-0.2cm}
\subsection{Normal order with respect to the true vacuum}\label{sec:normal_order}
\vspace{-0.2cm}

To this point, we have considered only normal order with respect to the Fermi vacuum. \pq also supports defining normal order relative to the true vacuum state, which can be useful for theories defined in terms of reduced density matrices (RDMs). One such method is the variational two-electron RDM (2RDM) approach,\cite{Percus64_1756,Rosina75_868,Rosina75_221,Garrod75_300,Rosina79_1366,Erdahl79_147,Fujisawa01_8282,Mazziotti01_042113,Mazziotti02_062511,Mazziotti06_032501,Percus04_2095,Zhao07_553,Lewin06_064101,Bultinck09_032508,DePrince16_423,DeBaerdemacker11_1235,VanNeck15_4064,DeBaerdemacker18_024105,Mazziotti17_084101,Ayers09_5558,Bultinck10_114113,Cooper11_054115,DePrince19_032509,Mazziotti08_134108,DePrince16_2260,Mazziotti16_153001} which involves the direct optimization of the elements of the 2RDM, subject to a subset of ensemble $N$-representability conditions.\cite{Coleman63_668} Many of these conditions involve enforcing correct mappings between RDMs. For example, when enforcing two-particle (PQG) conditions,\cite{Percus64_1756} a positive semidefinite 2RDM (${}^2${\bf D}) should map to positive semidefinite two-hole (${}^2${\bf Q}) and one-particle-one-hole (${}^2${\bf G}) RDMs.  The following code generates these mappings
\begin{lstlisting}[language=Python]
pq = pdaggerq.pq_helper('true')

# Q-matrix mapping
pq.set_string(['i', 'j', 'k*', 'l*'])
pq.add_new_string()
pq.simplify()
pq.print()
pq.clear()

# G-matrix mapping
pq.set_string(['i*', 'j', 'k*', 'l'])
pq.add_new_string()
pq.simplify()
pq.print()
pq.clear()
\end{lstlisting}
Note that we have introduced two new commands here. \texttt{set\_string} allows the user to set an arbitrary string of creation and annihilation operators that will be brought to normal order upon invokation of \texttt{add\_new\_string}. This code yields
\begin{lstlisting}[language=Python]
// normal-ordered strings:
//     + 1.00000 d(j,k) d(i,l) 
//     - 1.00000 l* i d(j,k) 
//     - 1.00000 d(i,k) d(j,l) 
//     + 1.00000 l* j d(i,k) 
//     + 1.00000 k* i d(j,l) 
//     - 1.00000 k* j d(i,l) 
//     + 1.00000 k* l* i j 


// normal-ordered strings:
//     + 1.00000 i* l d(j,k) 
//     - 1.00000 i* k* j l 
\end{lstlisting}
where \texttt{d(p,q)} represents a Kronecker delta function.
Higher order conditions such as the $(2, k)$-conditions\cite{Mazziotti12_062507} where $k \geq 3$ can be non-trivial to generate by hand, but with \pq translating these constraints to human readable expressions is {\color{black}easy}. For example, the following code generates mappings relevant to the T2 partial three-particle $N$-representability condition\cite{Erdahl78_697,Percus04_2095}
\begin{lstlisting}[language=Python]
pq = pdaggerq.pq_helper('true')

# T2 mapping
pq = pdaggerq.pq_helper('true')
pq.set_string(['i*','j*','k','n*','m', 'l'])
pq.add_new_string()
pq.set_string(['n*','m','l', 'i*','j*', 'k'])
pq.add_new_string()
pq.simplify()
pq.print()
pq.clear()
\end{lstlisting}
and yields
\begin{lstlisting}[language=Python]
// normal-ordered strings:
//     - 1.00000 i* j* l m d(k,n) 
//     + 1.00000 n* k d(i,l) d(j,m) 
//     - 1.00000 j* n* k m d(i,l) 
//     - 1.00000 n* k d(i,m) d(j,l) 
//     + 1.00000 j* n* k l d(i,m) 
//     + 1.00000 i* n* k m d(j,l) 
//     - 1.00000 i* n* k l d(j,m) 
\end{lstlisting}

\vspace{-0.8cm}
\section{Code Generation}
\vspace{-0.2cm}

\label{SEC:CODE_GENERATION}

An important part of computer algebra systems is the potential to translate equations into usable code.  \pq comes with a front-end parser that produces code for residual equations in a \np~einsum convention.  Appropriate summation limits, occupied or virtual indices, are enforced using \np~array slicing.  Using the \np~einsum format allows a user to implement the generated code with any of a variety of einsum backends.  For example, tensor contraction engines in Tensorflow~\cite{tensorflow2015-whitepaper}, Jax~\cite{jax2018github}, or Dask~\cite{matthew_rocklin-proc-scipy-2015} can be used to implement coupled cluster iterations on a variety of different hardware platforms.

Naive translation of contractions produced by \pq leads to correct, though inefficient, code.  For example, directly translating the doubles residual of the CCSD amplitude equations yields a tensor contraction that scale as $\mathcal{O}(n^{8})$ where $n$ is the number of spin-orbitals.  It is well known that the use of intermediate tensors and finding an optimal contraction ordering reduces the doubles residual scaling of CCSD to $\mathcal{O}(n^{6})$~\cite{Bartlett91_4334}.  The code generated by \pq's parser uses an exhaustive search of tensor contraction orderings provided by the \np~ einsum's \texttt{optimize=optimal} flag which was originally devised in the package of opt\_einsum~\cite{daniel2018opt}.  For example, the einsum line for a contraction required to implement the doubles residual is shown below. The code optimally contracts the antisymmeterized two electron integrals $g$ with two $t_{1}$ tensors and the $t_{2}$ tensor.
 \begin{lstlisting}[language=Python]
#	  0.5000 <i,j||a,b>*t1(e,i)*t1(f,j)*t2(a,b,m,n)
#   Complete contraction:  ijab,ei,fj,abmn->efmn
#          Naive scaling:  8
#      Optimized scaling:  6
# --------------------------------------------------
# scaling           current               remaining
# --------------------------------------------------
#    5        ei,ijab->abej      fj,abmn,abej->efmn
#    5        abej,fj->abef         abmn,abef->efmn
#    6      abef,abmn->efmn              efmn->efmn

doubles_res += 0.5 * einsum('ijab,ei,fj,abmn->efmn',
    g[o, o, v, v], t1, t1, t2, optimize=['einsum_path', 
    (0, 1), (0, 2), (0, 1)])
\end{lstlisting}  
\vspace{-0.2cm}
{\color{black} The parser also handles the permutation operator notation generated during simplification of contracted strings by introducing intermediate contracted tensors.  Examples of this can be seen in the code generation section of Appendix~\ref{app:ccsd_equation_gen}.}

The einsum code construction is in the Python component of \texttt{p$^\dagger$q}. Normal ordered strings or tensor contractions are parsed from a string based representation and translated into a Python object representation. Using the function \texttt{fully\_contracted\_stings()} associated with a \texttt{pq\_helper} object as input to \texttt{contracted\_strings\_to\_tensor\_terms()}, an internal representation of the contracted string is generated and stored. The stored objects are termed \texttt{TensorTerms} and have methods for simply generating the einsum contraction. These routines can be made aware of which labels are not summation indices (the \texttt{output\_variables} input field) which allows the same routine to be used for energy expressions, amplitude residual equations, etc. The following is an annotated example that produces code corresponding to the singles residual equations for CCSD.  
\begin{widetext}

\vspace{-0.2cm}
\begin{lstlisting}[language=Python]
import pdaggerq

pq = pdaggerq.pq_helper("fermi")  # initialize vacuum to Fermi vacuum

pq.set_left_operators(['e1(m,e)'])  # single excitation manifold

print('\t < 0 | m* e e(-T) H e(T) | 0> :')

pq.add_st_operator(1.0, ['f'], ['t1', 't2'])  # similarity transform the Fock operator with T1 and T2 generators
pq.add_st_operator(1.0, ['v'], ['t1', 't2'])  # similarity transform the fluctuation operator 
pq.simplify()

# grab list of fully-contracted strings, then print
singles_residual_terms = pq.fully_contracted_strings()
singles_residual_terms = contracted_strings_to_tensor_terms(singles_residual_terms) # convert to TensorTerms
for my_term in singles_residual_terms:
    print("#\t", my_term)
    print(my_term.einsum_string(update_val='singles_res', output_variables=('e', 'm'))) # generates einsum code
    print()
pq.clear()  # clear operators
\end{lstlisting}

\end{widetext}
which yields
\begin{lstlisting}[language=Python]
    < 0 | m* e e(-T) H e(T) | 0> :

#	  1.0000 f(e,m)
singles_res += 1.0 * einsum('em->em', f[v, o])

#	 -1.0000 f(i,m)*t1(e,i)
singles_res += -1.0 * einsum('im,ei->em', f[o, o], t1)

#	  1.0000 f(e,a)*t1(a,m)
singles_res += 1.0 * einsum('ea,am->em', f[v, v], t1)

#	  1.0000 f(i,a)*t2(a,e,i,m)
singles_res += 1.0 * einsum('ia,aeim->em', f[o, v], t2)

#	 -1.0000 f(i,a)*t1(a,m)*t1(e,i)
singles_res += -1.0 * einsum('ia,am,ei->em', f[o, v], t1, t1, optimize=['einsum_path', (0, 1), (0, 1)])

... elided output
\end{lstlisting}
This output code and similar code generated for the doubles residual equations can be copied into functions comprising a full CCSD iteration.  In Appendix~\ref{app:ccsd_equation_gen} we provide a full script for producing the energy, singles, and doubles contractions and in Ref.~\citenum{pq_2020} we provide a full implementation of the CCSD cluster amplitude equations, the CCSD de-excitation amplitude equations, and the equations for CCSD unrelaxed one- and two-particle RDMs.

Vacuum normal ordered strings can also be parsed into \texttt{TensorTerms} using the function
{\small\texttt{vacuum\_normal\_ordered\_strings\_to\_tensor\_terms}}.  This parsing function interprets the collection of normal ordered strings as RDM elements and Kronecker delta functions.  The resulting list of \texttt{TensorTerm} objects can be further simplified, rearranged, or augmented for any automatic code generation purpose.  For example, the matrix elements in extended RPA\cite{Rowe68_153,Pernal12_204109} theories can be represented as contractions of the Hamiltonian operator coefficient tensors and RDMs. Consider the extended RPA equations,
\begin{align}\label{eq:rpa_eq}
\sum_{ij}c_{ij}^{n} E_{ij,kl} = \omega_{n} \sum_{ij}c_{ij}^{n} S_{ij,kl},
\end{align}
with the $E_{ij,kl}$ and $S_{ij,kl}$ defined by
\begin{align}
    E_{ij,kl} = \langle \psi \vert \left[ a_{k}^{\dagger}a_{l}  \left[H, a_{j}^{\dagger}a_{i}\right]\right]\vert \psi \rangle\\ 
    S_{ij,kl} = \langle \psi \vert \left[a_{k}^{\dagger}a_{l}, a_{j}^{\dagger}a_{i}\right]\vert \psi \rangle
\end{align}
where $|\psi\rangle$ is the (possibly correlated) ground-state wave function. {\color{black}Note that here, unlike above, all labels refer to general orbitals ({\em i.e.}, $i$, $j$, $k$, and $l$ are not limited to the occupied space of any reference configuration)}. These matrix elements are straightforward to generate with \texttt{p$^\dagger$q}.  For $E_{ij,kl}$, for example, we simply generate an appropriate double commutator (See Tables \ref{tab:operators} and \ref{tab:functions}) and then use the \texttt{TensorTerm} einsum infrastructure to generate contraction code over the operator coefficients of the Hamiltonian. In this case we actually consider contractions over the reduced Hamiltonian elements, ${}^{2}K_{pqrs}$, defined by
\begin{align}
H =& \sum_{pqrs}{}^{2}K_{pqrs}a_{p}^{\dagger}a_{q}^{\dagger}a_{r}a_{s} \\
{}^{2}K_{pqrs} =& \frac{1}{4}\bigg (   \frac{1}{N-1} \left( h_{ps}\delta_{qr} - h_{qs}\delta_{pr} - h_{pr}\delta_{qs} \right. \nonumber \\
+& \left. h_{qr}\delta_{ps}\right) + \langle p q||s r\rangle \bigg )
\end{align}
where $N$ is the number of electrons. The following code
\begin{widetext}
\begin{lstlisting}[language=Python]
import pdaggerq

pq = pdaggerq.pq_helper('true') # normal order with respect to the true vacuum

# [k^l, [H, j^ i]] = - [[H,j^ i],k^l]
pq.add_double_commutator(-1.0,['e2(p,q,r,s)'],['e1(j,i)'],['e1(k,l)'])
pq.simplify()
rpa_tensor_terms = vacuum_normal_ordered_strings_to_tensor_terms(pq.strings())
pq.clear()

k2_idx = [Index('p', 'all'), Index('q', 'all'), Index('r', 'all'), Index('s', 'all')]
for tt in rpa_tensor_terms:
    # add the hamiltonian to contract with
    tt.base_terms += (pdaggerq.BaseTerm(indices=tuple(k2_idx), name='k2'),)
    print("\n# ", tt)
    print(tt.einsum_string(update_val='erpa_val',
                           occupied=['i', 'j', 'k', 'l', 'r', 's', 'p', 'q'],
                           virtual=[],
                           output_variables=['i','j','k','l'],
                           optimize=False))
\end{lstlisting}
yields
\begin{lstlisting}[language=Python]
#   1.0000 d(j,s)*d(i,k)*d2(p,q,l,r)*k2(p,q,r,s)
erpa_val += 1.0 * einsum('js,ik,pqlr,pqrs->ijkl', kd[:, :], kd[:, :], d2[:, :, :, :], k2[:, :, :, :])

#   1.0000 d(j,s)*d(k,r)*d2(p,q,i,l)*k2(p,q,r,s)
erpa_val += 1.0 * einsum('js,kr,pqil,pqrs->ijkl', kd[:, :], kd[:, :], d2[:, :, :, :], k2[:, :, :, :])

... elided output
\end{lstlisting}
\end{widetext}
The code generated above produces a four-tensor which can be reshaped into the matrix $E_{ij,kl}$.  The reduced Hamiltonian matrix elements and $2$-RDM are stored in OpenFermion order--$\langle 1'2'2 1 \rangle$.  Combining this result with the $1$-RDM based metric ($S_{ij,kl}$) to define the generalized eigenvalue problem that is the extended RPA equations.  Further simplifications can be made to resolve the delta functions but are left here, represented by identity matrices \texttt{kd[:, :]}, for clarity. 
Though further modifications can be applied, such as delta function simplification or taking advantage of antisymmetry of $K_{2}$, the resulting code is fully functioning computational kernel ready for deployment in an appropriate solver.
\section{Conclusions}
\label{SEC:CONCLUSIONS}

\pq augments the growing number of Python libraries to aid in fermionic many-body simulation and research. By implementing normal ordering and contraction routines along with algebraic simplifications in C++ and providing a high level interface with Python, the library seamlessly integrates with other Python based simulation tools while maintaining a high level of performance. \texttt{p$^\dagger$q}'s symbolic fermionic computer algebra system offers a broad set of functionality including bringing arbitrary fermionic algebra elements into normal order with respect to a Fermi or true vacuum, evaluating high order commutators with particular Hamiltonian elements, and similarity transforms generated by cluster operators up to {\color{black} fourth} order.   To ensure that this framework can aid in a variety of research areas related to fermionic many-body theories we provide functionality to generate tensor contraction code in the \np~einsum format that optimizes contraction order whenever possible. {\color{black} To demonstrate some of these features, we provide codes on GitHub\cite{pq_2020} that evaluate 1- and 2-RDMs at the CCSD level of theory, including the optimization of the underlying CCSD $t$- and $\lambda$-amplitudes. We also illustrate the ease with which high-order many-body approaches can be realized with working CCSD(T), CC3, CCSDT, and CCSDTQ solvers.}



Computer algebra systems have made computational implementation of many-body theories routine and lowered the rate of implementation errors.  \pq compliments prior works by providing a user friendly C++ accelerated Python variant geared towards rapid prototyping and extensibility.  While the first release of the library has focused mainly on code generation for {\color{black}spin-orbital formulations of} coupled cluster, more examples and utilities for applying \pq outside of this regime {\color{black}({\em e.g.}, spin  restriction)} are planned developments.  All code can be found on GitHub\cite{pq_2020} and is freely available licensed under Apache 2.0. {\color{black}Bug reports, feature requests, and community contributions are encouraged and can be made through GitHub's issue and pull request features.}

\section*{Acknowledgments}

This material is based upon work supported by the National Science Foundation under Grant No.~CHE-2100984.

\bibliography{main}

\onecolumngrid
\newpage
\appendix
\section{Python code generating tensor contractions for the CCSD residual equations}\label{app:ccsd_equation_gen}

The following code will generate \np~einsum contractions for the CCSD singles and doubles residual equations. A fully working spin-orbital CCSD code can be found at https://github.com/edeprince3/pdaggerq/. 

\begin{lstlisting}[language=Python]
import pdaggerq

from pdaggerq.parser import contracted_strings_to_tensor_terms

pq = pdaggerq.pq_helper("fermi")
pq.set_print_level(0)

# energy equation
pq.set_bra("")
print('\n', '    < 0 | e(-T) H e(T) | 0> :', '\n')
pq.add_st_operator(1.0, ['f'], ['t1', 't2'])
pq.add_st_operator(1.0, ['v'], ['t1', 't2'])

pq.simplify()

# grab list of fully-contracted strings, then print
energy_terms = pq.fully_contracted_strings()
pq.clear()
for my_term in energy_terms:
    print(my_term)
energy_terms = contracted_strings_to_tensor_terms(energy_terms)

for my_term in energy_terms:
    print("#\t", my_term)
    print(my_term.einsum_string(update_val='energy'))
    print()

# singles equations
pq.set_left_operators(['e1(m,e)'])
print('\n', '    < 0 | m* e e(-T) H e(T) | 0> :', '\n')
pq.add_st_operator(1.0, ['f'], ['t1', 't2'])
pq.add_st_operator(1.0, ['v'], ['t1', 't2'])
pq.simplify()

# grab list of fully-contracted strings, then print
singles_residual_terms = pq.fully_contracted_strings()
singles_residual_terms = contracted_strings_to_tensor_terms(singles_residual_terms)
for my_term in singles_residual_terms:
    print("#\t", my_term)
    print(my_term.einsum_string(update_val='singles_res', output_variables=('e', 'm')))
    print()
pq.clear()

# doubles equations
pq.set_left_operators(['e2(m,n,f,e)'])
print('\n', '    < 0 | m* n* f e e(-T) H e(T) | 0> :', '\n')
pq.add_st_operator(1.0, ['f'], ['t1', 't2'])
pq.add_st_operator(1.0, ['v'], ['t1', 't2'])
pq.simplify()

# grab list of fully-contracted strings, then print
doubles_residual_terms = pq.fully_contracted_strings()
doubles_residual_terms = contracted_strings_to_tensor_terms(doubles_residual_terms)
for my_term in doubles_residual_terms:
    print("#\t", my_term)
    print(my_term.einsum_string(update_val='doubles_res', output_variables=('e', 'f', 'm', 'n')))
    print()

pq.clear()
\end{lstlisting}

\newpage
\section{Python code for generating tensor contractions for the CCSD $\Lambda$-amplitudes equations}\label{app:lambda_equations}

The following code will generate \np~einsum contractions for the CCSD singles and doubles de-excitation residual equations.  A fully working spin-orbital $\Lambda$-CCSD code can be found at https://github.com/edeprince3/pdaggerq/.

\begin{lstlisting}[language=Python]
#  <0| e(-T) H e*m e(T)|0>
print('\n', '    0 = <0| e(-T) H e*m e(T)|0> + <0| L e(-T) [H, e*m] e(T)|0>', '\n')
pq.set_left_operators(['1'])
pq.set_right_operators(['1'])
pq.add_st_operator(1.0,['f','e1(e,m)'],['t1','t2'])
pq.add_st_operator(1.0,['v','e1(e,m)'],['t1','t2'])

# <0| L e(-T) [H,e*m] e(T)|0>
pq.set_left_operators(['l1','l2'])
pq.add_st_operator( 1.0,['f','e1(e,m)'],['t1','t2'])
pq.add_st_operator( 1.0,['v','e1(e,m)'],['t1','t2'])
pq.add_st_operator(-1.0,['e1(e,m)','f'],['t1','t2'])
pq.add_st_operator(-1.0,['e1(e,m)','v'],['t1','t2'])
pq.simplify()

# grab list of fully-contracted strings, then print
singles_residual_terms = pq.fully_contracted_strings()
singles_residual_terms = contracted_strings_to_tensor_terms(singles_residual_terms)
for my_term in singles_residual_terms:
    print("#\t", my_term)
    print(my_term.einsum_string(update_val='lambda_one', output_variables=('m', 'e')))
pq.clear()


#  <0| e(-T) H e*f*nm e(T)|0>
print('\n','    0 = <0| e(-T) H e*f*nm e(T)|0> + <0| L e(-T) [H, e*f*nm] e(T)|0>', '\n')
pq.set_left_operators(['1'])
pq.set_right_operators(['1'])
pq.add_st_operator(1.0,['f','e2(e,f,n,m)'],['t1','t2'])
pq.add_st_operator(1.0,['v','e2(e,f,n,m)'],['t1','t2'])

# <0| L e(-T) [H,e*f*nm] e(T)|0>
pq.set_left_operators(['l1','l2'])
pq.add_st_operator( 1.0,['f','e2(e,f,n,m)'],['t1','t2'])
pq.add_st_operator( 1.0,['v','e2(e,f,n,m)'],['t1','t2'])
pq.add_st_operator(-1.0,['e2(e,f,n,m)','f'],['t1','t2'])
pq.add_st_operator(-1.0,['e2(e,f,n,m)','v'],['t1','t2'])
pq.simplify()

# grab list of fully-contracted strings, then print
doubles_residual_terms = pq.fully_contracted_strings()
doubles_residual_terms = contracted_strings_to_tensor_terms(doubles_residual_terms)
for my_term in doubles_residual_terms:
    print("#\t", my_term)
    print(my_term.einsum_string(update_val='lambda_two', output_variables=('e', 'f', 'm', 'n')))
    print()
pq.clear()
\end{lstlisting}

\newpage
\section{Python code for generating equations for the CCSD one-particle RDM}
\label{app:1rdm}

The following code will generate equations for evaluating the elements of the one-particle RDM at the CCSD level of theory. A fully working code to evaluate the one-particle RDM can be found at https://github.com/edeprince3/pdaggerq/.

\begin{lstlisting}[language=Python]

import pdaggerq

pq = pdaggerq.pq_helper("fermi")

# D1(p,q) = <0|(1 + l1 + l2) e(-T) p*q e(T) |0> 
pq.set_left_operators(['1','l1','l2'])

print('\n', '    D1(m,n):', '\n')
pq.add_st_operator(1.0,['e1(m,n)'],['t1','t2'])
pq.simplify()
d1_terms = pq.fully_contracted_strings()
for my_term in d1_terms:
    print(my_term)
pq.clear()

print('\n', '    D1(e,f):', '\n')
pq.add_st_operator(1.0,['e1(e,f)'],['t1','t2'])
pq.simplify()
d1_terms = pq.fully_contracted_strings()
for my_term in d1_terms:
    print(my_term)
pq.clear()

print('\n', '    D1(e,m):', '\n')
pq.add_st_operator(1.0,['e1(e,m)'],['t1','t2'])
pq.simplify()
d1_terms = pq.fully_contracted_strings()
for my_term in d1_terms:
    print(my_term)
pq.clear()

print('\n', '    D1(m,e):', '\n')
pq.add_st_operator(1.0,['e1(m,e)'],['t1','t2'])
pq.simplify()
d1_terms = pq.fully_contracted_strings()
for my_term in d1_terms:
    print(my_term)
pq.clear()

\end{lstlisting}

\newpage
\section{Python code for generating equations for the CCSD two-particle RDM}
\label{app:2rdm}

The following code will generate equations for evaluating the elements of the two-particle RDM at the CCSD level of theory. A fully working code to evaluate the two-particle RDM can be found at https://github.com/edeprince3/pdaggerq/.

\begin{lstlisting}[language=Python]
import pdaggerq
pq = pdaggerq.pq_helper("fermi")

# D2(p,q,r,s) = <0|(1 + l1 + l2) e(-T) p*q*sr e(T) |0>
pq.set_left_operators(['1','l1','l2'])

blocks = [['i','j','k','l'],
          ['i','j','k','a'],
          ['i','j','a','l'],
          ['i','a','k','l'],
          ['a','j','k','l'],
          ['a','b','c','d'],
          ['a','b','c','i'],
          ['a','b','i','d'],
          ['i','b','c','d'],
          ['a','i','c','d'],
          ['i','j','a','b'],
          ['a','b','i','j'],
          ['i','a','j','b'],
          ['a','i','j','b'],
          ['i','a','b','j'],
          ['a','i','b','j']]
          
for my_block in blocks:
    print()
    print('#    D2(%s,%s,%s,%s)' %
         (my_block[0],
          my_block[1],
          my_block[2],
          my_block[3]))
    print()
    my_op = 'e2(' + my_block[0] + ',' \
                  + my_block[1] + ',' \
                  + my_block[3] + ',' \
                  + my_block[2] + ')'
    pq.add_st_operator(1.0,[my_op],['t1','t2'])
    pq.simplify()
    d2_terms = pq.fully_contracted_strings()
    for my_term in d2_terms:
        print("#\t", my_term)
    pq.clear()
\end{lstlisting}

\end{document}